\documentclass[12pt]{article}  
  
\usepackage{epsfig,epsf}  
\usepackage{graphics}  
\usepackage{cite}  
\usepackage{a4wide}  
\usepackage{xcolor}

\usepackage{amsmath}  
\usepackage{amsthm}  
\usepackage{amsfonts}  
\usepackage{amssymb}  
\usepackage{slashed}  
  
\renewcommand{\thefootnote}{\fnsymbol{footnote}}  
  
\newcommand{\be}{\begin{equation}}  
\newcommand{\ee}{\end{equation}}  
\newcommand{\ba}{\begin{eqnarray}}  
\newcommand{\ea}{\end{eqnarray}}  
\newcommand{\baa}{\begin{eqnarray*}}  
\newcommand{\btab}{\begin{tabular}}  
\newcommand{\etab}{\end{tabular}}  
\newcommand{\eaa}{\end{eqnarray*}}  
  
\newcommand \ket [1] {|{#1}\rangle}  
\newcommand \bra [1] {\langle {#1}|}

\numberwithin{equation}{section}  

\newcommand{\braket}[1]{\langle #1 \rangle}
\newcommand{\Tprod}[1]{{\mathrm T}\lbrack #1 \rbrack}

\newcommand{\note}[1]{{\small\bf\color{red}[#1]}}

\begin{document}  
  
\allowdisplaybreaks  
\thispagestyle{empty}  
  
\begin{flushright}  
{\small  
TUM-HEP-1002/15\\
June 19, 2015   
}  
\end{flushright}  
  
\vskip1.5cm  
\begin{center}  
\textbf{\Large\boldmath Higgs effects in top anti-top 
production near\\[0.1cm] threshold in $e^+ e^-$ annihilation}  
\\  
\vspace{1.2cm}  
{\sc M.~Beneke}$^a$, {\sc A.~Maier}$^a$, {\sc J. Piclum}$^b$
and {\sc T.~Rauh}$^a$  
\\[0.5cm]  
\vspace*{0.1cm} $^a$\,{\it 
Physik Department T31,\\ 
James-Franck-Stra\ss{}e~1,\\
Technische Universit\"at M\"unchen,\\
85748 Garching, Germany}\\[0.3cm]   
$^b$\,{\it 
Albert Einstein Center for Fundamental Physics,\\
Institute for Theoretical Physics,\\
Sidlerstrasse 5,\\
CH-3012 Bern, Switzerland}   
  
\def\thefootnote{\arabic{footnote}}  
\setcounter{footnote}{0}  
  
\vskip2cm  
\textbf{Abstract}\\  
\vspace{1\baselineskip}  
\parbox{0.9\textwidth}{
The completion of the third-order QCD corrections to the inclusive 
top-pair production cross section near threshold demonstrates that 
the strong dynamics is under control at the few percent level. In 
this paper we consider the effects of the Higgs boson on the cross 
section and, for the first time, combine the third-order QCD result 
with the third-order P-wave, the leading QED and the leading 
non-resonant contributions. We study the size of the different 
effects and investigate the sensitivity of the cross section 
to variations of the top-quark Yukawa coupling due to possible 
new physics effects.
}  
  
\end{center}  
  
  
\newpage  
\setcounter{page}{1} 
  
 
\section{Introduction}  

Top anti-top quark production near threshold in $e^+ e^-$ collisions 
provides a unique opportunity to measure the top-quark mass precisely, 
due to the well-defined center-of-mass energy and the enhancement 
of the cross section due to the strong-interaction Coulomb force. 
Whether the required theoretical precision on the cross section 
can be achieved has been an open question, since the 
second order (non-relativistic) QCD calculations revealed 
unexpectedly large corrections and uncertainties 
\cite{Beneke:1999qg,Hoang:2000yr}. After many years of work, the 
third-order QCD calculation has been recently finished \cite{BKMPPS}, 
resulting in a largely reduced theoretical uncertainty. With QCD 
effects under control, the emphasis shifts to other effects 
which must be addressed for a realistic cross section prediction. 
The most important are Higgs effects associated with the top-quark 
Yukawa coupling, general electromagnetic and electroweak 
corrections, non-resonant production of the final state $W^+ W^- b\bar b$ 
in the center-of-mass region near twice the top-quark mass, $2 m_t$, 
and photon initial-state radiation.

In this paper, we mainly focus on Higgs-boson effects and the 
sensitivity to the top-quark Yukawa coupling. The Yukawa potential 
generated by 
Higgs exchange~\cite{Strassler:1990nw} and one-loop corrections 
to $t\bar {t}$ production \cite{Guth:1991ab} have been considered 
long ago, but these early calculations do not reach the precision 
that corresponds to the third-order QCD calculation in the 
non-relativistic power-counting scheme. Third-order Higgs corrections 
to the production vertex and the energy and wave-function at the 
origin of a hypothetical S-wave toponium resonance have been 
computed in \cite{Eiras:2006xm}, but the $t\bar{t}$ cross section 
has not yet been considered. We supply this missing piece here.  
We also add for the first time the P-wave 
contributions \cite{Beneke:2013kia} and the leading non-resonant 
contributions \cite{Beneke:2010mp,Penin:2011gg} to the 
third-order S-wave QCD calculation. We then allow the top-quark 
Yukawa coupling $y_t$ to deviate from the Standard Model relation\footnote{
The symbol $v$ is used for the Higgs vacuum expectation value and 
the top-quark velocity, see below. The meaning should be clear from 
the context.} 
 $m_t = y_t v/\sqrt{2}$ and investigate the sensitivity to such 
deviations given the current theoretical uncertainties. 

  
\section{Higgs effects at NNNLO}  
\label{definition:Higgs}  
 
The contribution of the Higgs boson to the top pair production cross 
section $e^+ e^- \to t\bar{t}$ introduces two new parameters, the Higgs 
mass $m_H$, and top-quark  Yukawa coupling $y_t$. To set up the calculation 
we have to fix their relation to $m_t$ and the strong and electroweak 
couplings, $\alpha_s$ and $\alpha_{\rm EW}$, to establish 
the power counting. Recall that it is customary to count $\alpha_s\sim v$ and 
$\alpha_{\rm EW}\sim \alpha_s^2$, where $v = [(\sqrt{s}-2 m_t)/m_t]^{1/2}$ is 
the small top-quark velocity. A contribution of order 
$\alpha_s^k$ (or, equivalently, $v^k$) according to this counting 
is called ``N$^k$LO'' or ``$k$th order''. We opt for counting 
$y_t^2 \sim \alpha_{\rm EW}\sim \alpha_s^2$ and $m_H \sim m_t$. 
Other options would be to count the top-Yukawa coupling like 
the strong coupling, $y_t^2 \sim \alpha_s$, or the Higgs mass 
$m_H\sim m_t v$, or both. Clearly, 
with $m_t \approx 173\,$GeV, $m_H \approx 125\,$GeV and $v\sim 1/10$, 
the counting $m_H\sim m_t$ is more appropriate. In the terminology 
of non-relativistic effective theory and the threshold expansion, 
the Higgs mass is of order of the hard scale, and not the potential scale, 
which has significant impact on the structure of the contributions.
On the other hand the counting of the coupling simply determines
at which orders in the expansion the Higgs contributions appear
and we will justify our choice below.

The effective field theory setup is described in detail 
in~\cite{Beneke:2013jia}.
We recall that the dominant S-wave production cross section is proportional
to the imaginary part of the spectral function of the vector current 
\begin{equation}
 \Pi^{(v)}(q^2)=\frac{3}{2m_t^2}c_v^2G(E)+\dots,
 \label{eq:Piv}
\end{equation}
where $c_v$ is the hard matching coefficient of the vector current, 
$E=\sqrt{s}-2 m_t$,  and $G(E)$ is the Green function in 
potential-nonrelativistic QCD (PNRQCD), i.e. the
propagator of a non-relativistic top anti-top pair. The Higgs contributions to 
$c_v$ are discussed in Section~\ref{sec:shortdistanceeffects}.
To compute the corrections to the Green function the Higgs contributions
to the PNRQCD Lagrangian have to be determined.
Counting $m_H\sim m_t$ implies that the Yukawa-potential 
$\exp(-m_H r)/r$ generated by Higgs exchange between the top quarks 
is replaced by the local interaction $\delta^{(3)}(\mathbf{r})/m_H^2$ 
as is apparent from the Higgs propagator $1/(\mathbf{q}^{2}+m_H^2)$ 
in momentum space, where $\mathbf{q}^{2}\sim m_t^2 v^2$ can be 
neglected (expanded) relative to $m_H^2$. On the other hand, 
with $m_H\sim m_t v$, both terms would have to be kept. The 
contribution to the momentum-space potential is therefore simply
\begin{equation}
 \delta_H V=-\frac{y_t^2}{2m_H^2}.
 \label{eq:delHpotential}
\end{equation}
We note that this is suppressed by $v^3$ with respect to the leading 
QCD Coulomb potential $\alpha_s/\mathbf{q}^{2}$, where one power of $v$ 
arises from the counting of the Yukawa coupling, and two powers from 
the relative factor $\mathbf{q}^{2}/m_H^2$. The Higgs-induced potential
is thus a NNNLO effect. The corresponding correction to the Green function 
$G(E)$ is computed in Section~\ref{sec:potentialcontributions}.

\begin{figure} 
\begin{center} 
\includegraphics[width=4cm]{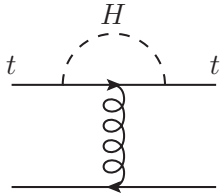}
\caption{One-loop Higgs correction to the colour Coulomb potential.}
\label{fig:deltaHVCoulomb}
\end{center} 
\end{figure}

Furthermore we have to consider corrections to the colour Coulomb
potential as shown in Fig.~\ref{fig:deltaHVCoulomb}. With $m_H\sim 
m_t$ counting, only the hard loop momentum region can yield a contribution. 
Since the external momenta are potential they have to be expanded, 
and we are left with an ${\cal O}(y_t^2)$ zero-momentum transfer correction 
to the $\psi^\dagger\psi A^0$ top-quark-gluon coupling of the NRQCD 
Lagrangian. However, since the top field is renormalized in the on-shell 
scheme this contribution cancels.

\subsection{Short-distance effects}  
\label{sec:shortdistanceeffects}
 
With $y_t\sim \alpha_s$ the leading, one-loop Higgs contribution 
to the hard matching coefficient 
of the vector current is of second order. It has been computed
in~\cite{Grzadkowski:1986pm,Guth:1991ab,Hoang:2006pd}.
Due to an additional diagram involving a $ZZH$ vertex, the matching
coefficients for the $\gamma t\bar{t}$ and the vector part of the
$Zt\bar{t}$ vertex differ. We neglect the contribution from
this diagram and use the $\gamma t\bar{t}$ coefficient, since the 
difference amounts to less than one percent of the already small NNLO Higgs 
contribution to the production vertex \cite{Guth:1991ab}. 
At NNNLO there are mixed
Higgs and QCD corrections to the vector current. They have been computed
as expansions for $m_H\approx m_t$ or $m_H\gg m_t$ in~\cite{Eiras:2006xm}, 
which is consistent with the adopted hard $m_H\sim m_t$ scaling. 
We use the result expanded in $(1-m_t^2/m_H^2)$, which was denoted
1$b$ in~\cite{Eiras:2006xm} and was shown to converge quickly for
Higgs boson masses around the physical value of about 125 GeV.
Based on the results of~\cite{Eiras:2006xm} we estimate the truncation
error due to this expansion to be well below one percent of the 
NNNLO matching coefficient and neglect it in the following.
The NNNLO correction to the hard matching contains an IR divergence,
which can be absorbed into a renormalization constant for the vector current
\begin{equation}
\widetilde{Z}_v=1+\left[\text{pure QCD}\right]+
\frac{\alpha_s C_F}{4\pi}\frac{y_t^2m_t^2}{2m_H^2}\frac{1}{4\epsilon}.
 \label{eq:Zv}
\end{equation}
The renormalized hard matching coefficient for the vector current can
be parametrized by
\begin{equation}
 c_v=1+\left[\text{pure QCD}\right]+\frac{y_t^2}{2}
\left[c_{vH}^{(2)}+\frac{\alpha_s}{4\pi}c_{vH}^{(3)}\right]+\dots,
 \label{eq:cvconv}
\end{equation}
where $c_{vH}^{(i)}$ can be obtained from~\cite{Eiras:2006xm}.
For convenience we reproduce the relevant expressions in the Appendix. 
The pure QCD correction is also known to NNNLO \cite{Marquard:2014pea}.
Starting at NNLO there is also a self-energy correction to the 
Z boson mediated cross section that contains the Higgs boson.
Since it does not involve the top Yukawa coupling we do not
consider it here. It will be added together with other NNLO electroweak
and non-resonant effects in future work.

To justify the power counting $y_t\sim\alpha_s$ we compare the
size of the Higgs effects discussed above to their QCD counterparts.
For the hard matching coefficient we obtain, with
$\alpha_s(\mu=80\text{ GeV})=0.1209$,
\begin{equation}
c_v=1-0.103|_{\alpha_s}-0.022|_{\alpha_s^2}+
0.031|_{y_t^2}-0.070|_{\alpha_s^3}-0.019|_{y_t^2\alpha_s}+\dots,
\label{Higgscv}
\end{equation}
where the contributions from different orders of the couplings
are shown explicitly. The power counting is clearly valid here.
For the potentials it is natural to compare the Higgs 
potential~\eqref{eq:delHpotential} to the spin-projected QCD NNLO Darwin 
potential $\delta V_D$, which is also local. Adopting $\alpha_s$ from above, 
we find $\delta_H V= -0.98/m_t^2$, which is only slightly smaller 
than $\delta V_D = {8\pi \alpha_s C_F}/(3 m_t^2) = 1.35/m_t^2$. 
However, since the Darwin potential yields only a small correction
compared to other NNLO effects we conclude that the overall 
counting is appropriate.

\subsection{Potential contributions}  
 \label{sec:potentialcontributions}
 
The potential correction to the Green function can be obtained
by quantum mechanical (PNRQCD) perturbation theory due to the instantaneous,
hence particle number conserving nature of potentials.
Since the Higgs potential \eqref{eq:delHpotential} is a NNNLO effect,
only the single insertion of $\delta_H V$ is required to compute the
NNNLO correction to the Green function
\begin{equation}
\delta_H G(E)=\bra{\mathbf{0}}\hat{G}_0(E)\,i\delta_H V\,i\hat{G}_0(E)
\ket{\mathbf{0}}=-\delta_H V\,G_0(E)^2.
\label{eq:deltaHGreen}
\end{equation}
The remarkably simple form arises because of the locality of
the potential. The Green function $G_0(E)$ describes the propagation
of a top quark pair, produced and destroyed at zero
spatial separation, under the influence of the leading-order 
QCD Coulomb potential. The insertion of a local interaction thus
factorizes into the product of a
Green function to the left and the right of the insertion.
Using the well-known result for the LO Green function
\begin{equation}
G_0(E)=\frac{m_t^2\alpha_s C_F}{4\pi}
\left[\frac{1}{4\epsilon}+L_\lambda+\frac12-\frac{1}{2\lambda}-
\hat\psi(1-\lambda)+\mathcal{O}(\epsilon)\right],
\label{eq:G0}
\end{equation}
in $d=4-2\epsilon$ space-time dimensions, 
expressed through $\lambda=\alpha_sC_F/(2\sqrt{-E/m_t})$,
$L_\lambda=\log(\lambda\mu/(m_t\alpha_sC_F))$, and 
$\hat\psi(x) = \gamma_E+\psi(x)$,
we observe that the imaginary part of \eqref{eq:deltaHGreen} is UV divergent,
\begin{equation}
 \left.\text{Im}\left[\delta_H G(E)\right]\right|_\text{div}=
\frac{y_t^2}{m_H^2}\frac{m_t^2\alpha_s C_F}{16\pi\epsilon}\,
\text{Im}\left[G_0(E)\right],
 \label{eq:divdeltaHGreen}
\end{equation}
where $G_0(E)$ denotes the exact $d$-dimensional LO Green function.
Note that this expression also contains a finite term from the $1/\epsilon$
pole multiplying the unknown $\mathcal{O}(\epsilon)$ part of the
LO Green function. However,~\eqref{eq:divdeltaHGreen} exactly cancels in the 
combination $\text{Im}[(c_v\widetilde{Z}_v^{-1})^2G(E)]$. 
This is analogous to the cancellation between the divergent part 
of the QCD Darwin potential single insertion and part of the 
divergence of the two-loop QCD contribution to $c_v$. In the following, 
we therefore only have to consider the finite part  
\begin{equation}
\delta_H G_\text{fin}(E) = \frac{y_t^2}{2m_H^2}\left(
\frac{m_t^2\alpha_s C_F}{4\pi}\left[L_\lambda+\frac12-\frac{1}{2\lambda}-
\hat\psi(1-\lambda)\right]\right)^2.
\label{eq:deltaHGreenfin}
\end{equation}

Due to the non-perturbative treatment of the LO
Coulomb potential in PNRQCD, the exact Green function contains
single poles below threshold, which correspond to $^3S_1$ toponium 
bound states: 
\begin{equation}
 G(E)\mathop{\longrightarrow}^{E\rightarrow E_n}
\frac{\left|\psi_n(0)\right|^2}{E_n-E-i\epsilon}.
\label{eq:GreenfunctionBS}
\end{equation}
The energy levels $E_n$ and squared wave functions at the origin, 
$|\psi_n(0)|^2$, also receive corrections from the insertion of the Higgs 
potential. We use the parametrization
\begin{equation}
\begin{aligned}
 E_n&=E_n^{(0)}\left(1+[\text{pure QCD}]+
\frac{\alpha_s}{4\pi}\frac{y_t^2}{2}e_H\right),\\
 |\psi_n(0)|^2&=|\psi_n^{(0)}(0)|^2\left(1+[\text{pure QCD}]+
\frac{\alpha_s}{4\pi}\frac{y_t^2}{2}f_H\right),
\end{aligned}
\end{equation}
where
\begin{equation}
 E_n^{(0)}=-m_t\left(\frac{\alpha_sC_F}{2n}\right)^2,
\hspace{2cm}
|\psi_n^{(0)}(0)|^2=\frac1\pi\left(\frac{m_t\alpha_sC_F}{2n}\right)^3.
\end{equation}
The corrections $e_H,f_H$ can be obtained by expanding 
\eqref{eq:deltaHGreenfin} and \eqref{eq:GreenfunctionBS}
around the bound-state energies and comparing coefficients.
Alternatively both equations can be expanded around positive
integer values of $\lambda$. For \eqref{eq:deltaHGreenfin}
we obtain
\begin{equation}
\delta_H G_\text{fin}(E)=\frac{y_t^2}{2m_H^2}
\frac{m_t^4\alpha_s^2C_F^2}{16\pi^2}
\left[\frac{1}{(n-\lambda)^2}+\frac{2}{n-\lambda}
\left(L_n+\frac12-\frac{1}{2n}-\hat\psi(n)\right)+\dots\right],
\label{eq:deltaHGreenfinexp}
\end{equation}
where $L_n=\log(n\mu/(m_t\alpha_sC_F))$ and the ellipsis denotes
terms that are regular in the limit $\lambda\rightarrow n$. We obtain
\begin{equation}
e_H=\frac{m_t^2C_F}{m_H^2}\frac2n,
\hspace{1cm}
f_H=\frac{m_t^2C_F}{m_H^2}\left(2L_n+1+\frac{4}{n}-2S_1(n)\right).
\end{equation}
Here $S_1(n)=\sum_{k=1}^nk^{-1}$ denotes the harmonic number of order one.
The result for $e_H$ agrees with \cite{Eiras:2006xm} and for $f_H$ we
reproduce the value for $n=1$ given in \cite{Eiras:2006xm}.

\subsection{Combined}

Combining hard and potential effects due to the Higgs boson, more  
precisely, the top-Yukawa interaction with the Higgs boson, we can express 
the NNLO correction to the vector correlation function (\ref{eq:Piv})
as 
\begin{equation}
 \delta_{2H}\Pi^{(v)}=\frac{3}{2m_t^2}y_t^2c_{vH}^{(2)}G_0(E).
 \label{eq:delta2HPi}
\end{equation}
The NNNLO correction is
\begin{equation}
\delta_{3H}\Pi^{(v)}=\frac{3}{2m_t^2}\left[
\frac{\alpha_s}{4\pi}y_t^2(c_{vH}^{(3)}+c_{vH}^{(2)}c_1)G_0(E)+
y_t^2c_{vH}^{(2)}\delta_1G(E)+\delta_H G_\text{fin}(E)\right].
\label{eq:delta3HPi}
\end{equation}
It includes cross terms of the NNLO Higgs correction with the known NLO QCD 
corrections to the Green function, $G(E) = G_0(E)+\delta_1G(E)+\ldots$, 
and matching coefficient $c_v=1+\alpha_s c_1/(4 \pi)+\ldots$. Both terms 
are finite, as it is understood here that $G_0(E)$ from (\ref{eq:G0}) 
is minimally subtracted. The physical cross section is related to 
the imaginary part of $\Pi^{(v)}(q^2)$.


\section{Non-resonant and QED effects} 
\label{sec:others} 
 
In the computation of QCD and Higgs corrections to the $t\bar{t}$ cross
section the top decay width has been accounted for by the replacement
$E\rightarrow E+i\Gamma_t$, where both quantities are of the
same order. Since the top quark is unstable, one should
rather consider the production cross section for the decay
product $W^+W^-b\bar{b}$ of the top pair.\footnote{The 
W boson can be treated as stable here, since its kinematics is not 
sensitive to the W width.} This final state can also
be produced without an intermediate resonant top pair and only the 
sum of both processes constitutes a physical quantity.
This is also apparent from an incomplete cancellation
of UV divergences and scale dependence in the non-relativistic 
description of the resonant process starting at NNLO. The 
non-resonant correction is important for realistic cross section 
predictions, since it affects particularly the cross section 
below the position of the peak, where the sensitivity to the 
top-quark mass is largest~\cite{BKMPPS}.

The computation of the $e^+e^-\rightarrow W^+W^-b\bar{b}$
cross section in the top anti-top threshold region, such that it is 
consistent with the non-relativistic power counting, 
expansion and resummation of the resonant sub-process, 
can be performed in the framework of 
unstable-particle effective theory \cite{Beneke:2003xh,Beneke:2004km}.
The master formula for the cross section is
\begin{eqnarray}
\label{eq:master}
i {\cal A} &=&\sum_{k,l} C^{(k)}_p  C^{(l)}_p \int d^4 x \,
\braket{e^- e^+ |
\Tprod{i {\cal O}_p^{(k)\dagger}(0)\,i{\cal O}_p^{(l)}(x)}|e^- e^+}
\nonumber\\ 
&& + \,\sum_{k} \,C_{4 e}^{(k)} 
\braket{e^- e^+|i {\cal O}_{4e}^{(k)}(0)|e^- e^+}.
\\[-0.5cm]
\nonumber
\end{eqnarray}
The first line describes the resonant production of the $t\bar t$ pair 
through an operator ${\cal O}_p^{(k)}$. Since the
initial state is colour-neutral, the matrix element further
factorizes into a leptonic and a hadronic tensor as long as
only QCD and no electroweak effects are considered, and the
non-relativistic treatment in terms of the  PNRQCD Green function 
$G(E)$ is recovered. The second line describes non-resonant 
production of the $W^+W^-b\bar{b}$ final state, which is why the hadronic 
contribution can be absorbed fully into a hard Wilson coefficient. 
The leading non-resonant effects appear already at NLO in the non-relativistic 
power counting, and have been determined in~\cite{Beneke:2010mp}, including 
the possibility of imposing invariant-mass cuts on the top decay 
products. The result for the total cross section was confirmed 
in \cite{Penin:2011gg} with an independent method. In the following analysis 
we combine the results from~\cite{Beneke:2010mp} with the third-order 
QCD calculation \cite{BKMPPS}. We exclude the small contribution from 
$e^+e^-\rightarrow W^+W^-H$ followed by $H\to b\bar b$, since 
it can be considered as a reducible ``background'' and eliminated by 
an invariant-mass cut on the $b\bar b$ jets as discussed 
in~\cite{Beneke:2010mp}. 
At NNLO only partial results for the non-resonant term in 
the second line of (\ref{eq:master}) are 
available~\cite{Penin:2011gg,Jantzen:2013gpa,Ruiz-Femenia:2014ava}.
We do not consider them here and hope to include the complete
NNLO non-resonant correction together with other NNLO electroweak effects
in future work. We note that the cancellation
of divergences at NNLO has already been demonstrated 
\cite{Jantzen:2013gpa} and that the contribution is likely
to be numerically relevant below threshold.

There is, however, one further electroweak effect already at NLO. 
The counting $\alpha \sim \alpha_s^2$ of the QED coupling implies 
that the QED Coulomb potential 
$\delta V_\text{QED}=-4\pi\alpha Q_t^2/\mathbf{q}^2$  
represents a NLO correction relative to the leading order QCD Coulomb 
potential. While only a single insertion of this potential would be 
required for NLO accuracy, we include contributions involving the NLO 
QED Coulomb potential up to NNNLO, i.e.~we also include multiple insertions 
of $\delta V_\text{QED}$ as well as mixed insertions together with 
other potentials and current matching coefficients. The required 
expressions can be inferred from the known results for multiple insertions 
of the QCD potentials \cite{Beneke:2005hg,paperII}, 
but one has to be careful when considering insertions which contain divergences
in the imaginary part, since the QCD Coulomb potential contains
$\mathcal{O}(\epsilon)$ terms which are absent in the QED potential.
The QED potential has already been included in previous 
calculations \cite{Pineda:2006ri,Hoang:2010gu,Beneke:2010mp}, 
but not yet in combination with the third-order QCD 
result \cite{BKMPPS}. Similarly, the numerical effect of 
the NLO non-resonant terms was studied in detail in \cite{Beneke:2010mp}, 
but was not implemented so far in the code that includes the 
higher-order QCD corrections. 
 

\section{Size of Higgs and other non-QCD effects}  
\label{sec:analysis}  

For the cross section predictions shown below we employ the 
values
\begin{equation}
m_t^{\rm PS}(20\,\mbox{GeV})=171.5\,\mbox{GeV}, \quad \Gamma_t=1.33\,\mbox{GeV}
\end{equation}
for the top-quark PS mass \cite{Beneke:1998rk} and top-quark width. 
We note that the QED Coulomb potential is not part of the definition of the
PS mass, since, in contrast to QCD, higher-order QED corrections do
not give rise to an IR renormalon ambiguity in the pole mass, and are 
rapidly decreasing. The strong and electromagnetic couplings are 
\begin{equation}
\alpha_s(M_Z) = 0.1185\pm 0.006,\quad 
\alpha(M_Z) = 1/128.944,
\label{eq:alpha}
\end{equation}
where the QCD coupling refers to the $\overline{\rm MS}$ scheme and the 
running QED coupling, taken from \cite{Hagiwara:2011af}, 
to the on-shell scheme.  
These parameters are taken to be consistent with \cite{BKMPPS}. We further 
use 
\begin{equation}
M_W= 80.385\,\mbox{GeV},\quad
M_Z= 91.1876\,\mbox{GeV},\quad
m_H =125\,\mbox{GeV}
\end{equation}
for the electroweak gauge and Higgs boson masses, from which we derive 
the Weinberg angle, Higgs expectation value $v$ and top-Yukawa coupling 
$y_t = \sqrt{2} m_t/v$ through tree-level relations. Here $m_t$ is the 
top pole mass, computed from the PS mass with NNNLO accuracy.

The QCD NNNLO result includes the P-wave contribution 
at the same order, computed in \cite{Beneke:2013kia}, which arises from 
production of the $t\bar t$ pair through the axial-vector coupling of 
a virtual $Z$-boson. This enhances the $S$-wave cross section presented in 
\cite{BKMPPS} by about 1\%. The theoretical uncertainty of the cross 
section calculation itself is estimated by varying the renormalization 
scale between 50 and 350~GeV. The ``default scale'' is set to 80~GeV. 
The ``finite-width'' factorization 
scale $\mu_w$ related to the separation of resonant and non-resonant terms 
in (\ref{eq:master}) is fixed to $\mu_w=350\,$GeV. The dependence on 
this scale is cancelled exactly order-by-order in the sum of the 
two contributions. Since presently the non-resonant terms 
are included only to NLO, while the resonant terms are known to NNNLO, 
there is a small uncancelled dependence on $\mu_w$ at NNLO. 

\begin{figure}[p] 
\begin{center} 
\includegraphics[width=11cm]{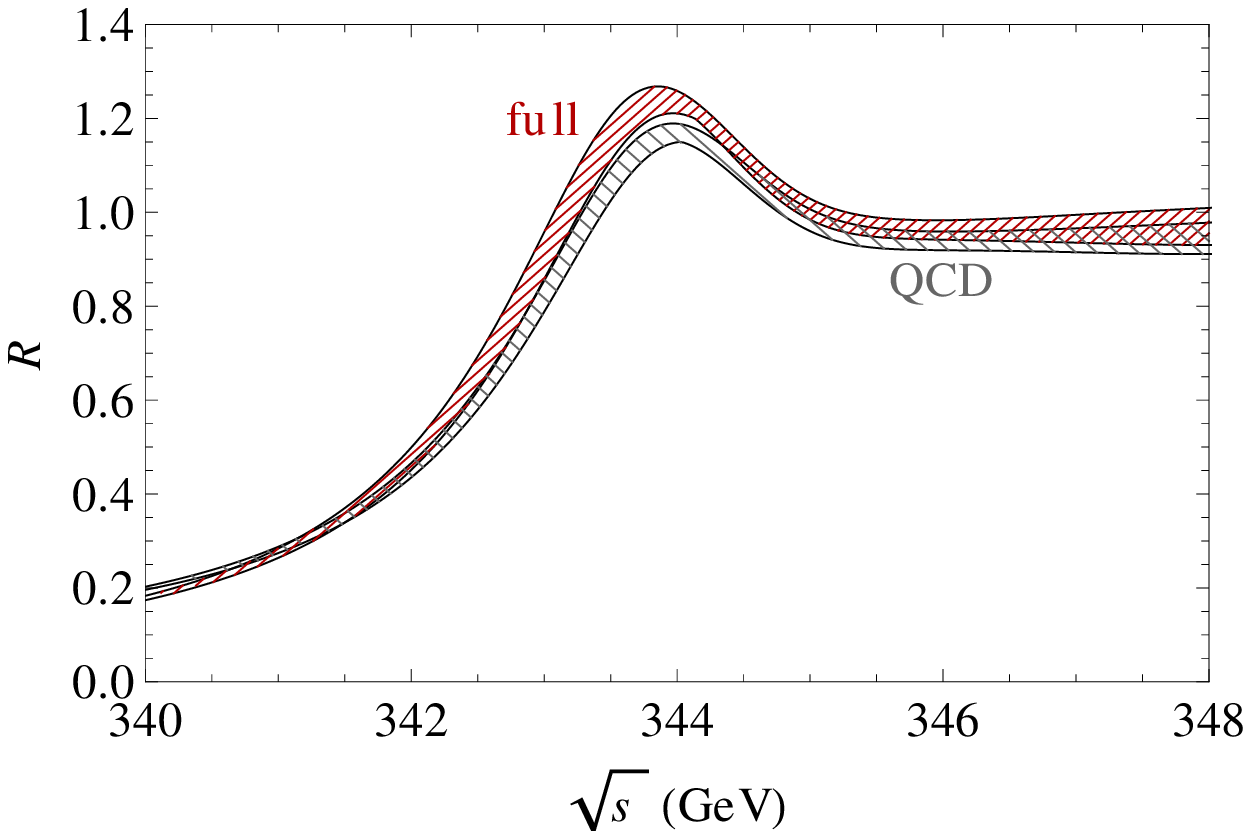}

\vspace*{0.5cm}
\hskip-0.3cm  
\includegraphics[width=11.4cm]{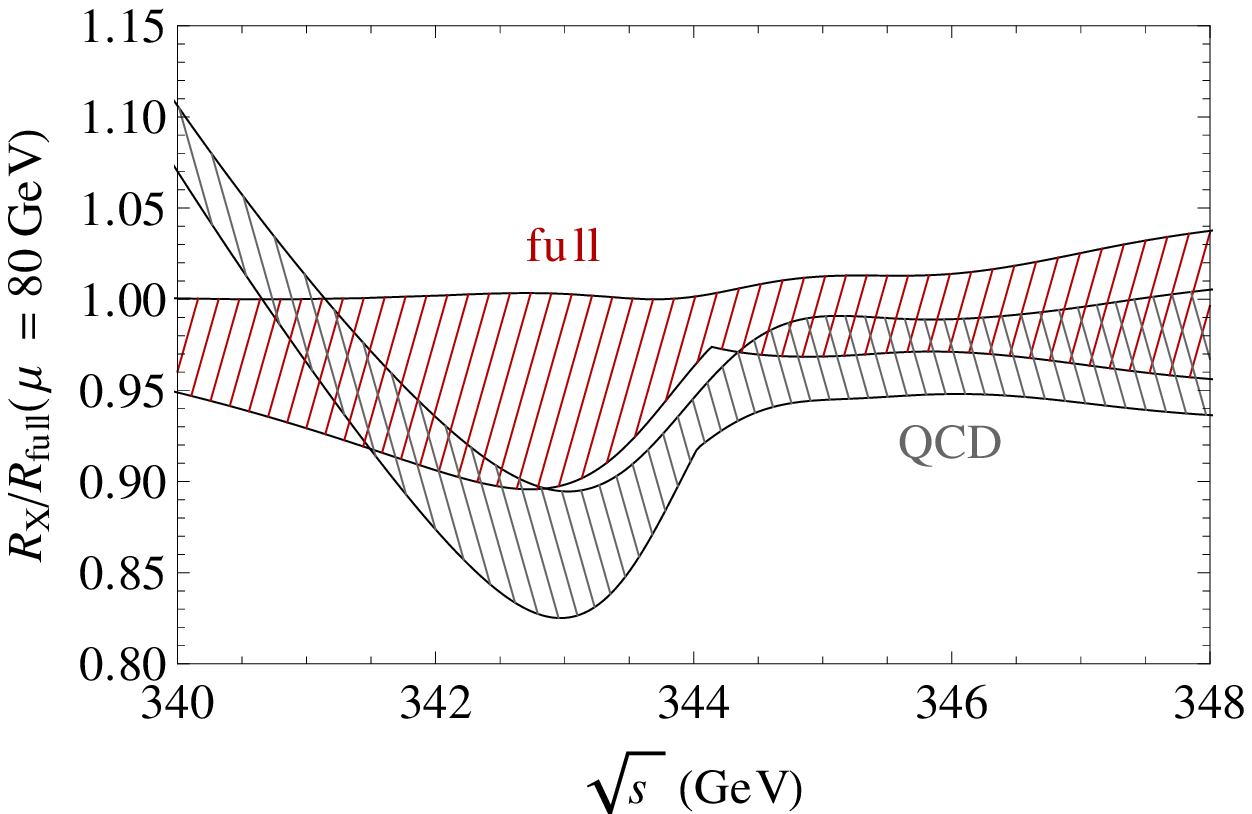} 
\end{center} 
\caption{The $R$ ratio for QCD-only (light-grey hatched band) and including
  the Higgs, QED, and non-resonant contributions (red/dark-grey hatched 
  band) as functions of the center-of-mass energy. The bands are due to the
  scale variation. The upper plot shows the absolute results and the 
  lower plot the results normalized to the full one
  evaluated at the scale $\mu=80$~GeV.}  
\label{fig:xsection}  
\end{figure}  

In Fig.~\ref{fig:xsection} we show the total $e^+ e^- \to 
W^+ W^- b\bar b$ cross section in the range of $e^+ e^-$ 
center-of-mass energy $\sqrt{s}$ a few GeV below and above the top anti-top 
threshold including 
the Higgs, QED and non-resonant corrections discussed above 
(red/dark-grey hatched band) and compare it to the QCD-only result at NNNLO 
(light-grey hatched band). The cross section is
always normalized to the LO cross section for $e^+e^- \to \mu^+\mu^-$,
yielding the so-called $R$ ratio $R=\sigma(e^+ e^- \to W^+
W^- b\bar b)/\sigma_0$, with $\sigma_0=4\pi\alpha^2/(3s)$.
The QED and Higgs potentials lead to an attractive force, which 
enhances the cross section. The leading Higgs 
contribution to the short-distance coefficient (\ref{Higgscv}) 
is also positive, resulting in an overall enhancement of 
about 10\% near and 5\% above the peak. Below the peak 
the negative non-resonant contribution becomes important and wins 
over the QED and Higgs enhancement. The lower plot in Fig.~\ref{fig:xsection} 
contains the same results as the upper one, but now all values have been 
normalized to the full $R$ ratio evaluated for $\mu=80$~GeV. Comparing 
the two bands, we again observe the enhancement of the full cross section
around and above the peak. The rise in the QCD-only result at energies
below the peak is due to the non-resonant
contribution, which decreases the full cross section in this
region. We observe a small increase in the scale uncertainty of the full 
result relative to QCD-only, which can now reach $\pm 5\%$ about one 
GeV below the peak, but is mostly of the $\pm 3\%$ size. The additional 
scale dependence arises mainly from the Higgs potential insertion.

\begin{figure}[t] 
\begin{center} 
\includegraphics[width=10cm]{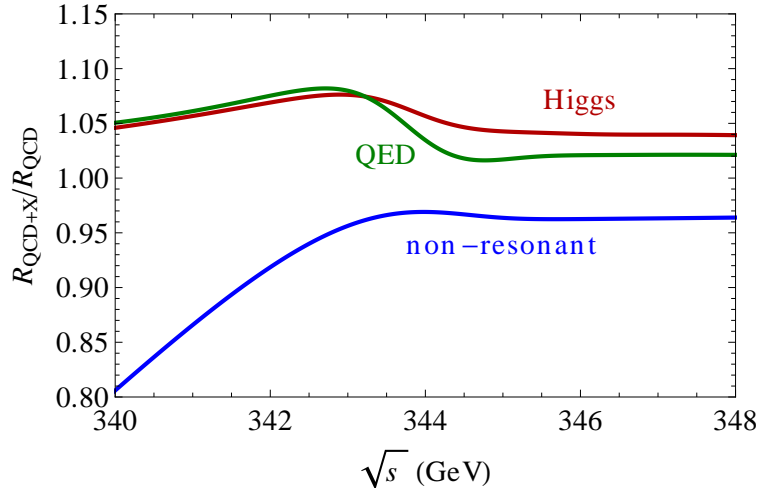}  
\end{center} 
\caption{The sum of the QCD and individual non-QCD contributions to
  the $R$ ratio, normalized to the QCD-only result, as functions of
  the center-of-mass energy.}
\label{fig:relativesize}  
\end{figure}  

\begin{figure}[p] 
\begin{center} 
\includegraphics[width=11cm]{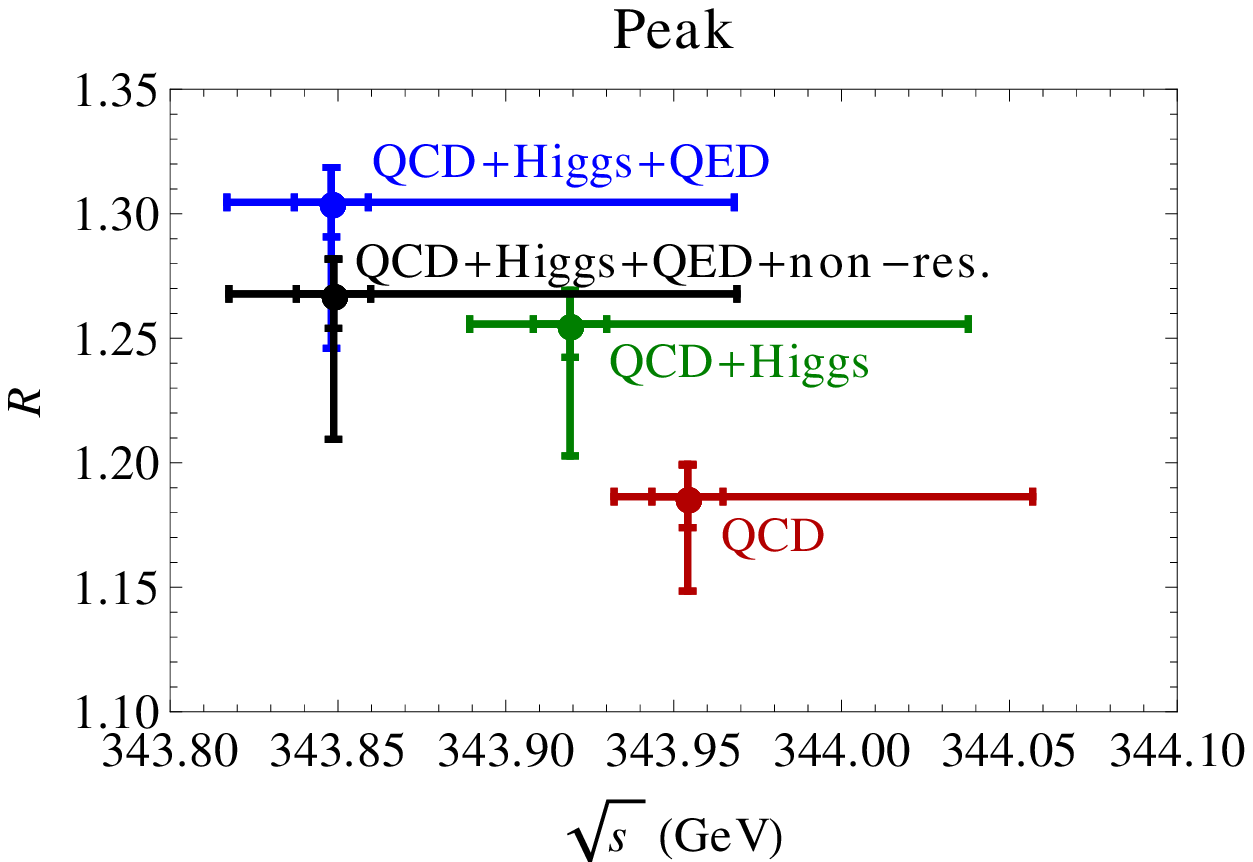}  
\vskip0.5cm
\includegraphics[width=11cm]{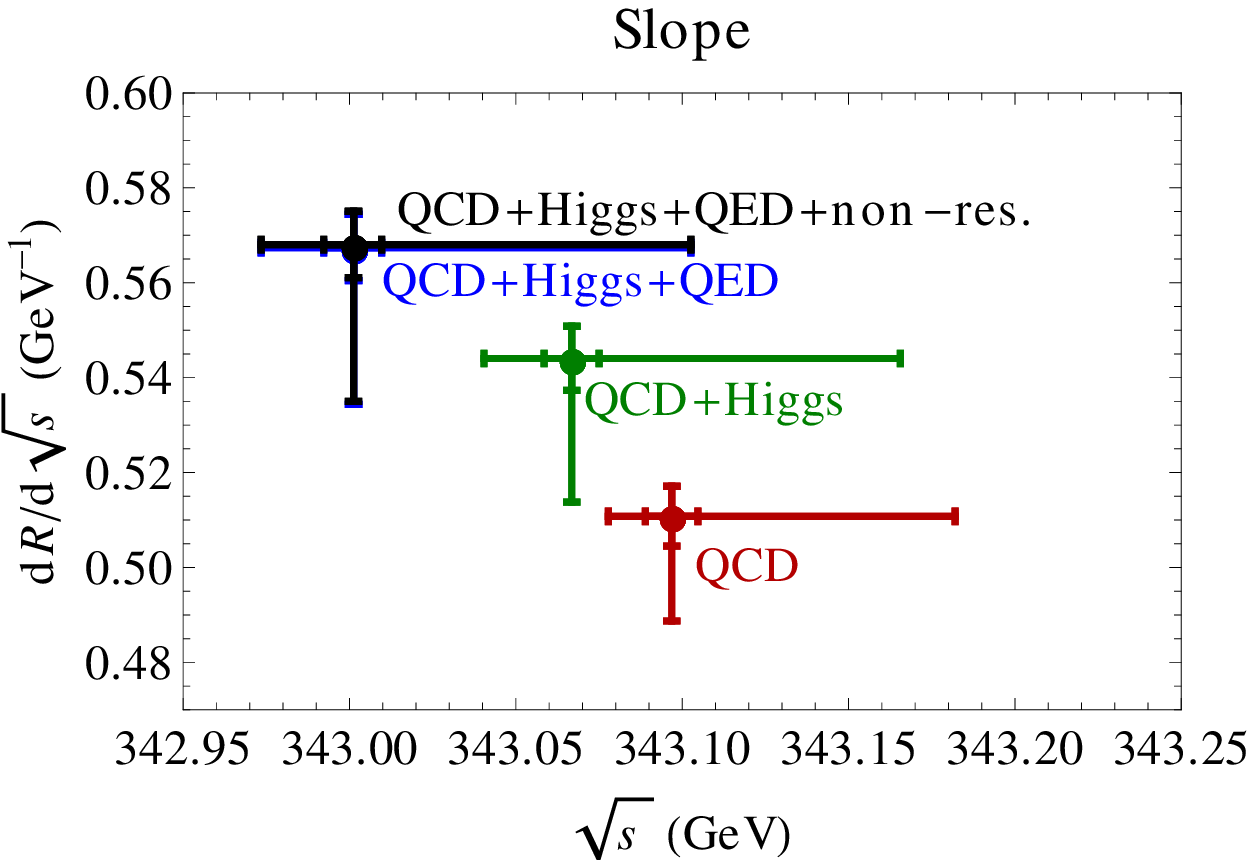}  
\end{center} 
\caption{The upper plot shows the peak height versus the peak position
  of the $R$ ratio for QCD-only (red), QCD+Higgs (green),
  QCD+Higgs+QED (blue) and including all contributions (black). The
  inner error bar denotes the uncertainty due to $\alpha_s(M_Z)$,
  while the outer one denotes the quadratic sum of the scale and
  $\alpha_s$ uncertainties. The lower plot is the same but for the
  maximal slope. In this case there is no visible change due to
  the non-resonant contribution.}  
\label{fig:peakslope}  
\end{figure}

The size of the three non-QCD contributions considered here--- the Higgs, 
QED and non-resonant contributions--- are shown separately in 
Fig.~\ref{fig:relativesize}, normalized 
to the NNNLO QCD-only result for the total $t\bar{t}$ cross section 
and plotted as function of the $e^+ e^-$ center-of-mass energy $\sqrt{s}$ 
(always $\mu=80\,$GeV).
The peak of the QCD-only cross section for the adopted parameters is at 
$\sqrt{s}=343.95\,$GeV (see Fig.~\ref{fig:peakslope}).
The Higgs and QED contributions result in positive shifts of the cross
section of about 4--8\% for the former and 2--8\% for the latter, depending 
on the value of $\sqrt{s}$. Both shift
the peak to lower energies, thus resulting in larger enhancements below the
peak of the QCD-only result. The NLO non-resonant correction 
is a nearly energy-independent, negative contribution~\cite{Beneke:2010mp} 
in absolute size and is therefore increasingly important below threshold, 
which explains the shape of the corresponding line in 
Fig.~\ref{fig:relativesize}.
Its absolute size is smaller than the sum of the QED and Higgs contributions 
in the peak region and above, which leads to the overall positive shift in 
these regions observed in Fig.~\ref{fig:xsection}. Below threshold the resonant
contribution falls off quickly and the relative correction from the 
non-resonant part becomes very large, up to 20\%. The same behaviour
is found for the dependence on the scale $\mu_w$. For variations in 
physically reasonable ranges from the potential to the hard scale we find 
a relative uncertainty of less than $\pm 1\%$ above threshold, but 
up to $\pm 3\%$ a few GeV below threshold. 
The dominant part of this uncertainty 
comes from the NNLO corrections and cancels exactly once the NNLO 
non-resonant corrections are included, after which the remaining 
$\mu_w$ dependence is at most 1\%. We note the uncertainty from 
$\mu_w$ variation is not included in Fig.~\ref{fig:xsection} and 
subsequent figures that show scale variations.

For a determination of the top-quark mass from the threshold cross section,
the sharp rise and the peak in the cross section are its most important
features. On the other hand, determinations of the top-quark decay width and
its Yukawa coupling require a precise knowledge of the overall
normalization. In order to judge the influence of the Higgs, QED, and
non-resonant contribution on these quantities, Fig.~\ref{fig:peakslope} shows
the impact on the peak (upper plot) and maximal slope (lower plot) of 
the cross section, when these
contributions are added successively to the QCD-only result. The theoretical
uncertainty due to the variation of $\alpha_s(M_Z)$ within its uncertainty
given in~(\ref{eq:alpha}) is shown as the inner error bars. The outer
error bars are the quadratic sum of the $\alpha_s$ and scale uncertainty. They
provide an indication of the significance of the changes.

The Higgs and QED contributions result in a negative shift of the peak
position and an increase in the peak height. Correspondingly, the position of
the maximal slope is also shifted to a lower energy and its value is increased
when these contributions are added to the QCD result. The Higgs contribution
shifts the peak position by $-35$~MeV and the QED contributions adds another
$-71$~MeV. Since the peak position is related to twice the top-quark mass,
this translates into a $-53$~MeV difference in the top-quark masses obtained
from the full and QCD-only results for the cross section. Since the
non-resonant contribution is an almost energy-independent negative shift, it
has almost no influence on the position of the peak and only decreases its
height. It also leaves the slope unchanged. Therefore, it is mostly important
for the overall normalization of the cross section.

\section{Sensitivity to the top Yukawa coupling}  
\label{sec:topyukawa}  

The mechanism of fermion mass generation in the Standard Model (SM) is 
intimately related to the question whether the Yukawa coupling of the 
Higgs boson to fermion $f$ is proportional to the fermion's mass, 
$y_f = \sqrt{2} m_f/v$. In the SM effective theory including dimension-six 
operators \cite{Buchmuller:1985jz,Grzadkowski:2010es} this relation 
can be violated, for example, by the operator 
\begin{equation}
\Delta\mathcal{L}=-\frac{c_\text{NP}}{\Lambda^2}
(\phi^\dagger\phi)(\bar{Q}_3\widetilde{\phi}t_R) + \rm{h.c.},
\end{equation}
where $c_\text{NP}$ is a new, independent coupling, $\Lambda$
the scale of new physics and $\widetilde{\phi}=i\sigma^2\phi^*$.
For simplicity we have neglected flavor indices
and assume the new physics to only affect the third generation.
After spontaneous symmetry breaking, the operator generates 
corrections to the top mass term and Higgs coupling, 
\begin{equation}
\Delta\mathcal{L}\supset -\frac{c_\text{NP}v^2}{2\sqrt{2}\Lambda^2}
\left(v \bar t_L t_R +3 h \bar t_L t_R\right) + \rm{h.c.},
\end{equation}
where $h$ denotes the physical Higgs field. 
We observe that the coefficients of the 
mass and Yukawa term differ, and obtain the relation
\begin{equation}
\kappa_t \equiv \frac{y_t}{\sqrt{2} m_t/v} 
= 1+ \frac{c_{\rm NP}}{\Lambda^2}\frac{v^3}{\sqrt{2} m_t},
\label{eq:Yukawamod}
\end{equation}
where
\begin{equation}
m_t=\frac{v}{\sqrt{2}}\left(y_t^\text{SM}+
\frac{c_\text{NP}v^2}{2\Lambda^2}\right)
\quad\mbox{and}\quad
y_t = y_t^\text{SM}+
\frac{3 c_\text{NP}v^2}{2\Lambda^2}.
\end{equation}
Below we will use $\kappa_t$ defined 
in~\eqref{eq:Yukawamod} to parametrize corrections to the standard
relation between $y_t$ and $m_t$. To investigate the sensitivity of the 
top anti-top cross section to $\kappa_t$, we do not use the 
SM relation between $m_t$ and $y_t$ in the calculation of the Higgs 
contribution, and rescale $y_t$ by $\kappa_t$ in the Higgs potential 
(\ref{eq:delHpotential}) and the short-distance contributions. That is, 
we simply assume that some new physics effect makes the top mass and Yukawa 
coupling independent parameters. Evidently, 
the complete set of dimension-six operators may induce further anomalous 
couplings of the top quark, such as an anomalous top-gluon coupling, 
which can give additional short-distance and potential contributions 
to the cross section. A full treatment of these effects is beyond 
the scope of this work. 

\begin{figure}[p] 
\begin{center} 
\includegraphics[width=10cm]{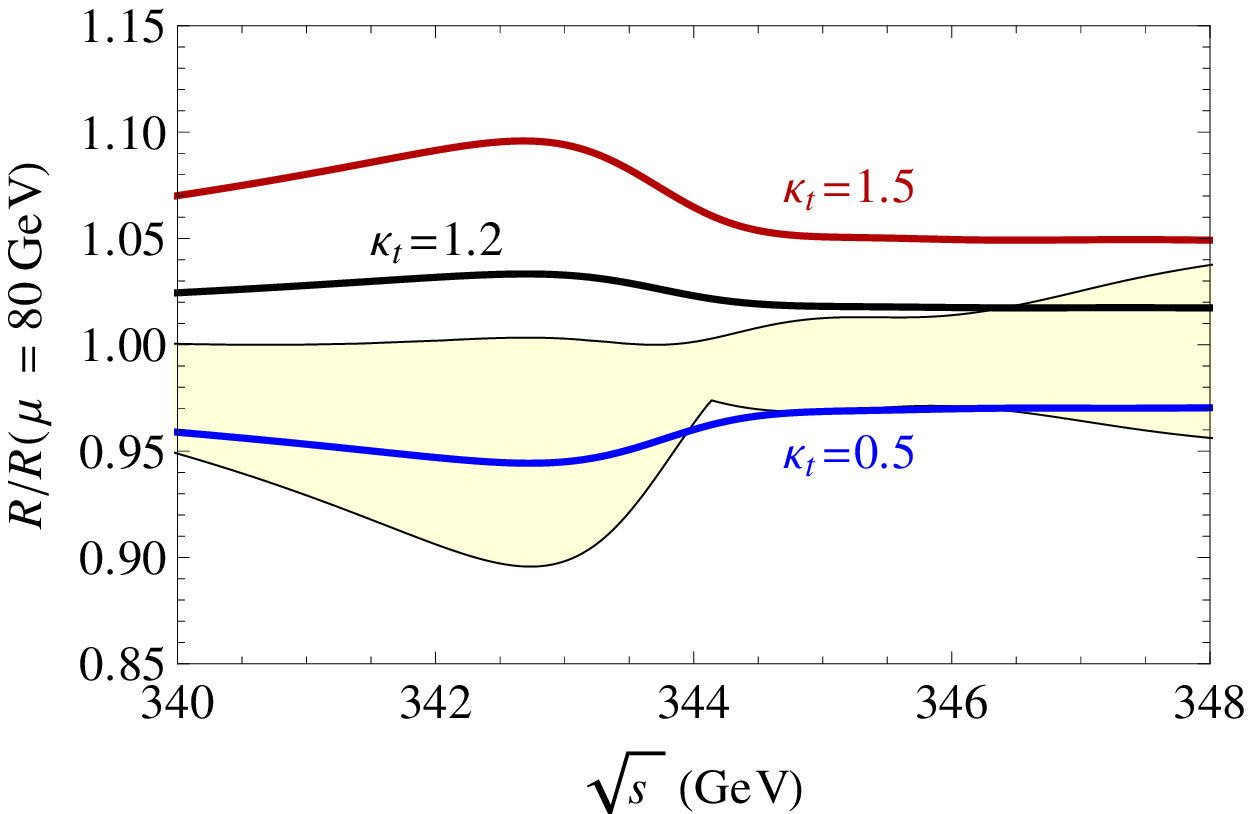}  

\vspace*{0.3cm}
\includegraphics[width=10cm]{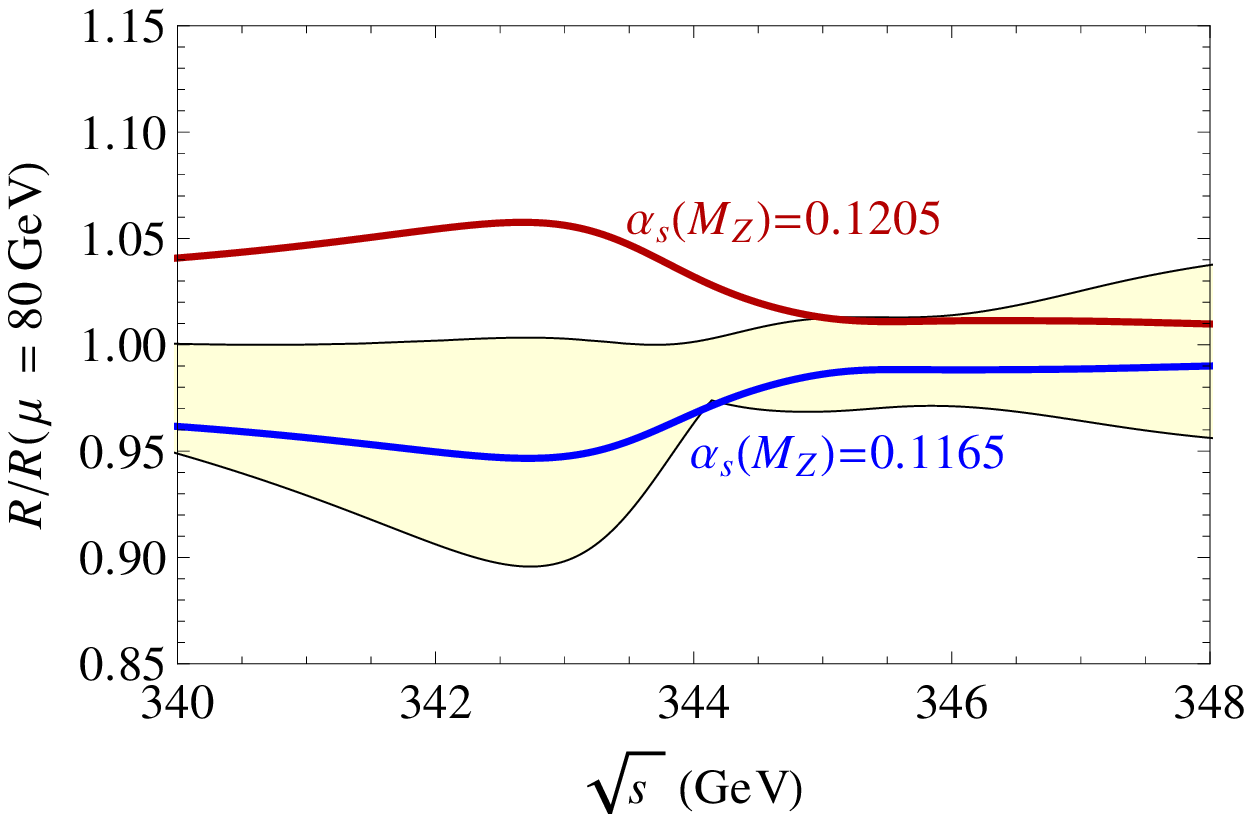}  
\end{center} 
\caption{The sensitivity of the $R$ ratio to the variation of the
  Yukawa coupling (upper plot) and the strong coupling (lower
  plot). The bands denote the uncertainty due to scale variation of
  the $R$ ratio for $y_t=y_t^{\mathrm{SM}}$ ($\kappa_t=1$) and
  $\alpha_s(M_Z)=0.1185$. All values have been normalized to the $R$
  ratio evaluated at $\mu=80$~GeV.}
\label{fig:yukasvar}  
\end{figure}  

The sensitivity of the $R$ ratio to variations of the Yukawa coupling
is shown in the upper plot of Fig.~\ref{fig:yukasvar}. The plot shows
curves for different values of $\kappa_t$ normalized to the result at 
$\kappa_t=1$ and $\mu=80$~GeV. The main effect of an
increase (decrease) in the Yukawa coupling is a strengthening
(weakening) of the attractive potential between the top and anti-top
quarks. This results in an increase of the cross section of 5--10\%
for $\kappa_t=1.5$ or a decrease of 3--5\% for $\kappa_t=0.5$, with some
dependence on the center-of-mass energy. In order to provide a first
estimate of the possible precision of a Yukawa coupling measurement
from top anti-top threshold production, the plot also shows the
theoretical uncertainty due to the variation of the renormalization scale
$\mu$. Naively one would expect to be only sensitive to values of the
Yukawa coupling that lie outside this uncertainty band. From the
figure we see that this requires rather large deviations from the SM
value of roughly $+20\%$ or $-50\%$. However, a more detailed analysis
should also take into account the shape of the curve, which may lead
to an improved sensitivity.

\begin{figure}[t] 
\begin{center} 
\includegraphics[width=11cm]{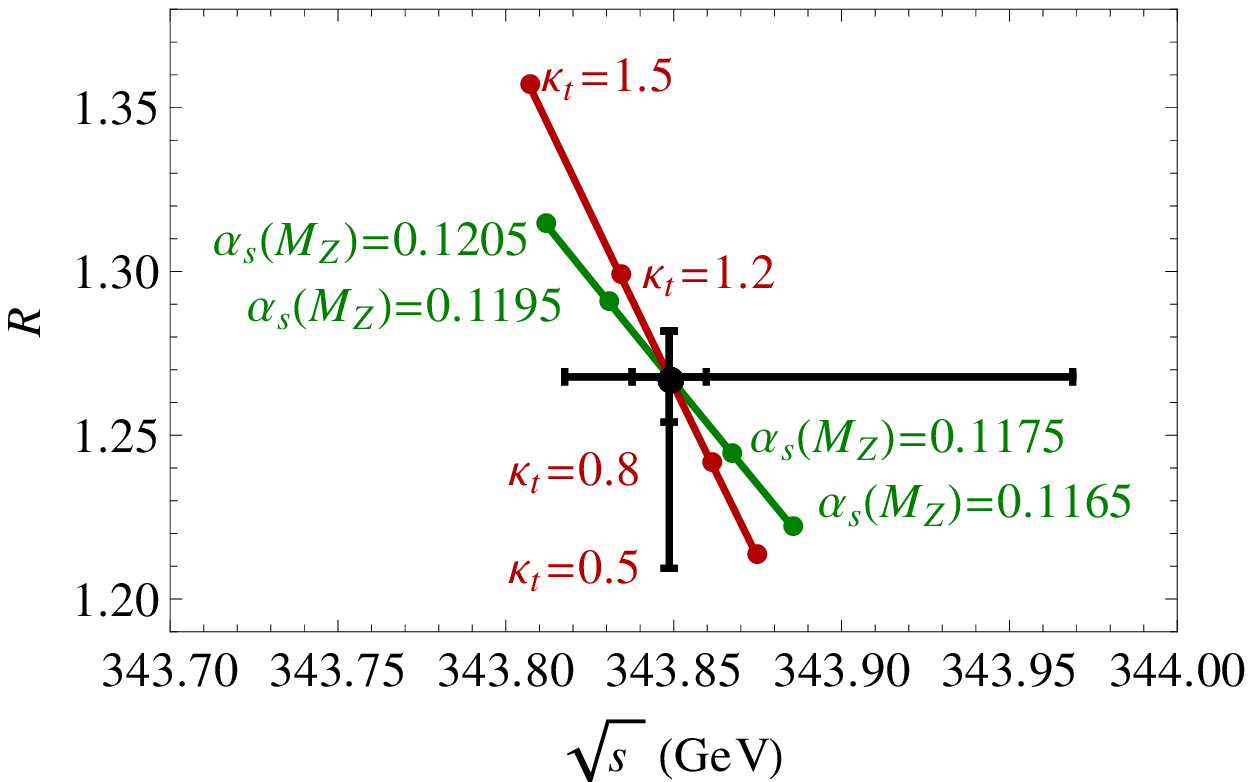}  
\end{center} 
\caption{Changes in peak height and position due to variation of the
  Yukawa coupling (red line) and the strong coupling (green line). The
  black error bars denote the $\alpha_s$ and combined scale and
  $\alpha_s$ uncertainty for $y_t=y_t^{\mathrm{SM}}$ ($\kappa_t=1$) and
  $\alpha_s(M_Z)=0.1185$ (cf. upper plot in
  Fig.~\ref{fig:peakslope}).}
\label{fig:yukaspeak}  
\end{figure}  

Another important point is that a variation of the strong coupling
leads to similar changes in the cross section as a variation of the
Yukawa coupling. This can be seen by comparing the upper and lower
plot in Fig.~\ref{fig:yukasvar}, where the lower one shows curves for
$\alpha_s(M_Z)=0.1205$ and $\alpha_s(M_Z)=0.1165$. Just as for the
Yukawa coupling, an increase (decrease) of the strong coupling leads
to an increased (decreased) cross section, though the energy
dependence of the shift is slightly different. To make this point
clearer, Fig.~\ref{fig:yukaspeak} shows the change in height and
position of the peak of the $R$ ratio due to changes in the strong and
Yukawa coupling. The similar slope of the resulting lines indicates
the degeneracy in the variations of the two
parameters. Thus, the precision of a Yukawa coupling measurement
depends on the uncertainty of $\alpha_s(M_Z)$.
Alternatively, one could perform a simultaneous fit of both
couplings, though this would again lead to a loss in precision for
the Yukawa coupling.

Ref.~\cite{Horiguchi:2013wra} finds that for a Higgs boson with a mass of
about 125~GeV the top quark Yukawa coupling can be obtained with a statistical
uncertainty of only 4.2\%. This result is based on the increase of the 
cross section from $y_t=0$ to $y_t=y_t^{\mathrm{SM}}$, which is
assumed to be 9\% and energy independent. Neither the
theoretical uncertainty of the cross section, nor the correlation with the
strong coupling constant are considered. Our results show that once 
theoretical uncertainties are taken into account, it is unlikely that 
such a high precision can be achieved.

\section{Conclusion}  
\label{conl:sec}  

The completion of the NNNLO QCD correction to the top anti-top production 
cross section near threshold has increased the precision of the theoretical 
prediction to a level where non-QCD effects gain importance. 
In this paper we added NNNLO Higgs, and the leading (NLO) QED and 
non-resonant contributions to the third-order QCD result 
for the top anti-top production cross section. All three effects are 
larger than the current QCD uncertainty of about $\pm 3\%$ 
\cite{BKMPPS} and cause a distinct modification of the cross section 
below, near and above the peak. We quantified the theoretical uncertainty 
in the presence of these effects, the dependence on the strong coupling, 
and the sensitivity to a modification
of the top-Yukawa coupling. Further studies 
should be performed in the framework of realistic simulations accounting 
for beam and initial-state radiation effects. On the theoretical side, 
the inclusion of NNLO electroweak and non-resonant corrections would 
further sharpen the prediction, especially below the resonance 
peak. 

\subsubsection*{Acknowledgements}
We thank Y.~Kiyo and P.~Ruiz-Femen\'ia for comments, and M.~Steinhauser 
for comments and helpful communications about Ref.~\cite{Eiras:2006xm}.
This work was supported by the Gottfried Wilhelm Leibniz programme 
of the Deutsche For\-schungs\-gemeinschaft (DFG)
and the DFG cluster of excellence ``Origin and Structure of the Universe".

\begin{appendix} 
\section{Higgs contribution to the hard matching coefficient} 
\label{appendix} 

For convenience we give the expressions used for the Higgs contribution 
to the hard matching coefficient of the vector current~\eqref{eq:cvconv}. 
The results are taken from~\cite{Eiras:2006xm}, only the prefactors have 
been adjusted to match our convention.
 
\begin{equation}
 c_{vH}^{(2)}=\frac{1}{\pi^2}\left[\frac{3z-1}{12z}-\frac{2-9z+12z^2}{48z^2}\ln z+\frac{2-5z+6z^2}{24z}\Psi(z)\right],
\end{equation}
where $z=m_t^2/m_H^2$ and
\begin{equation}
\Psi(z)=\begin{cases}
\displaystyle \,\frac{\sqrt{4z-1}}{z}\,
\text{arctan}\sqrt{4z-1},\hspace{1cm}&z\geq1/4,\\[0.5cm]
\displaystyle \,\frac{\sqrt{1-4z}}{2z}
\ln\frac{1-\sqrt{1-4z}}{1+\sqrt{1-4z}},\hspace{1cm}&z<1/4.
\end{cases}
\end{equation}
\begin{eqnarray}
  c_{vH}^{(3)}&=&\frac{4C_F}{\pi^2}\Big[\frac{\pi^2}{8}(1-y)\ln\frac{m_t^2}{\mu^2}-5.760+5.533y-0.171y^2+0.0124y^3+0.0304y^4
    \nonumber\\
  &&\phantom{\frac{4C_F}{\pi^2}}+0.0296y^5+\dots\Big],
\end{eqnarray}
where $y=1-z$.

\end{appendix}  
  
\bibliographystyle{JHEP-2}

\providecommand{\href}[2]{#2}\begingroup\raggedright\endgroup


\end{document}